\documentclass[10pt, final]{IEEEtran}
\usepackage{amsthm}

\usepackage{amssymb}
\usepackage{amsmath}
\usepackage{threeparttable}
\usepackage{bbm}
\usepackage[linesnumbered,ruled,vlined]{algorithm2e}
\usepackage{mathrsfs}
\usepackage{cite}
\usepackage{bm,epsfig,amsthm,url}
\usepackage{indentfirst}
\usepackage{amsfonts}
\usepackage{epstopdf}
\usepackage{cases}
\usepackage[caption=false,font=scriptsize]{subfig}
\usepackage{xcolor}
\usepackage{mathtools}

\makeatletter
\renewcommand{\maketag@@@}[1]{\hbox{\m@th\normalsize\normalfont#1}}%
\makeatother
\usepackage{hyperref}
\usepackage{stfloats}
\usepackage{algpseudocode}
\usepackage{booktabs}
\usepackage{hyperref}

\newtheoremstyle{mystyle}{}{}{}{}{}{: }{0pt}{\indent \it{\thmname{#1}\thmnumber{ #2}\thmnote{#3}}}
\theoremstyle{mystyle}

\newtheorem{Proposition}{Proposition}

\usepackage[letterpaper, left=0.68in, right=0.68in, bottom=1in, top=0.72in]{geometry}
\def\BibTeX{{\rm B\kern-.05em{\sc i\kern-.025em b}\kern-.08em
    T\kern-.1667em\lower.7ex\hbox{E}\kern-.125emX}}
\allowdisplaybreaks[4]

\begin{document}

\title{Joint Sparsity and Beamforming Design for RDARS-Aided Systems}
\author{Chengwang~Ji, Haiquan~Lu, Qiaoyan~Peng, Jintao~Wang, and~Shaodan~Ma
\thanks{
C. Ji, Q. Peng, J. Wang, and S. Ma are with the State Key Laboratory of Internet of Things for Smart City and the Department of Electrical and Computer Engineering, University of Macau, Macao SAR, China (e-mails: ji.chengwang@connect.um.edu.mo, \{qiaoyan.peng, wang.jintao\}@connect.um.edu.mo; shaodanma@um.edu.mo).
H. Lu is with the School of Electronic and Optical Engineering, Nanjing University of Science and Technology, Nanjing 210094, China (e-mail: haiquanlu@njust.edu.cn).}
\vspace{-30pt}
}
\maketitle
\begin{abstract}
Reconfigurable distributed antennas and reflecting surface (RDARS) has emerged as a promising architecture for communication performance enhancement.
In particular, the new selection gain can be achieved by leveraging the dynamic working mode selection between connection and reflection modes, whereas low-complexity element configuration remains an open issue.
In this paper, we consider a RDARS-assisted communication system, where the connected elements are formed as a uniform sparse array for simplified mode configuration.
The sum rate maximization problem is then formulated by jointly optimizing the active and passive beamforming matrices and sparsity of connected element array.
For the special cases of a single user equipment (UE) and two UEs, the optimal sparsity designs are derived in closed-form. Then, for an arbitrary number of UEs, a weighted minimum mean-square error-based alternating optimization (AO) algorithm is proposed to tackle the non-convex optimization problem. Numerical results demonstrate the effectiveness of low-complexity sparsity optimization.
\end{abstract}

\begin{IEEEkeywords}
Reconfigurable distributed antennas and reflecting surface (RDARS), uniform sparse array, mode switching, joint sparsity and beamforming design.
\end{IEEEkeywords}
\vspace{-10pt}
\section{Introduction}
With the explosive growth of wireless data traffic in the sixth-generation (6G) communications, several promising technologies, such as extremely large-scale multiple-input multiple-output (XL-MIMO) \cite{haiquan_survey}, distributed antenna system (DAS) \cite{larsson2024massive}, and reconfigurable intelligent surface (RIS) \cite{wu_RIS, chen2025joint}, have been proposed to meet the ambitious terabit-per-second peak data requirement \cite{haiquan_survey, xue_survey}.
Recently, the reconfigurable distributed antennas and reflecting surface (RDARS) has been proposed, which integrates the benefits of the DAS and RIS \cite{ChengzhiMa_ANewArchi, Wang_RDARS, ji2025model, zhang_RDARS, ji2025reconfigurable}, while avoiding the active noise issue suffered by active RIS.
Moreover, RDARS includes RIS and flexible-position multi-state RIS \cite{ren2025flexible} as special cases. 
Specifically, the working mode of each element can be flexibly switched between the connection mode and reflection mode via a controller \cite{ChengzhiMa_ANewArchi}, thus yielding an additional selection gain compared to passive RIS and multi-state RIS \cite{Wang_RDARS, zhang_RDARS, chen2021hybrid}.

To explore the potential of RDARS, the theoretical performance was first analyzed under the fixed positions of RDARS connected elements \cite{ChengzhiMa_ANewArchi}, where the distributed and reflection gains were revealed compared to the DAS and RIS-aided system. Subsequently, by exploiting the flexible working mode switching, the selection gain was investigated to further improve the system performance \cite{ji2025reconfigurable, Wang_RDARS, zhang_RDARS, ji2025model}. In \cite{ji2025reconfigurable}, the reconfigurable codebook was proposed to jointly design beamforming and placement positions of RDARS elements.
However, the codebook-based beamforming design and working mode selection incur the performance loss due to quantization errors. Consequently, optimization-based beamforming and mode switching methods have been adopted to overcome this issue. 
For example, in \cite{Wang_RDARS}, a block coordinate descent-based penalty dual decomposition algorithm was proposed to minimize the total mean-square-error (MSE).
where the binary constraint was equivalently transformed into a more tractable formulation and solved by the majorization-minimization (MM) technique. 
In \cite{zhang_RDARS}, the maximization of radar output signal-to-noise ratio (SNR) was investigated by jointly designing the beamforming and mode selection.
where the mode selection was reformulated as a sorting problem.
Nevertheless, the optimization of mode selection matrix suffers from the practical issue of high computational complexity, such as $\mathcal{O}(N^{3.5})$ in \cite{Wang_RDARS} and $\mathcal{O}(9N^{3})$ in \cite{zhang_RDARS}, where $N$ denotes the total number of RDARS elements, which may limit the practical applications. 

Recently, sparse arrays have attracted growing research interest, where the inter-element spacing is typically larger than the half wavelength as in the conventional compact arrays, and a larger physical aperture can be achieved given the identical number of connected elements \cite{wang2023can, lu2025flexible}. Then, a higher spatial degree of freedom (DoF) comes for enhanced communication and sensing performance \cite{lu2025flexible}. 
As such, an interesting idea is to place the RDARS connected elements in a sparse array for simplified mode configuration.
However, how to characterize the performance in terms of the sparsity level of connected element array (CEA) remains unknown, where the sparsity level refers to the separation of adjacent elements. Moreover, how to determine the sparsity of CEA is unclear for practical scenarios.

To fill this gap, in this paper, we investigate a RDARS-aided multi-user system, where the CEA is arranged as a uniform sparse array. The sum rate is maximized by jointly optimizing the active and passive beamforming, as well as the sparsity of CEA. To obtain useful insights, we first consider two special cases of the single-UE and two-UE.
For the single-UE case, we derive the SNR expression, and the result shows that it is independent of the spatial distribution of CEA. For the two-UE case, the closed-form expressions of the sparsity level to minimize the channel’s squared-correlation coefficient (CSCC) are derived under different channel conditions.
Moreover, for an arbitrary number of UEs, we propose an efficient weighted minimum mean-square error (WMMSE)-based
alternating optimization (WA) algorithm to tackle the non-convex optimization problem.
Numerical results verify the effectiveness of the low-complexity sparsity design under different number of UEs.

\section{System Model and Problem Formulation} \label{sec: system model}
\begin{figure}[t]
 \centering
  \includegraphics[width=0.35\textwidth]{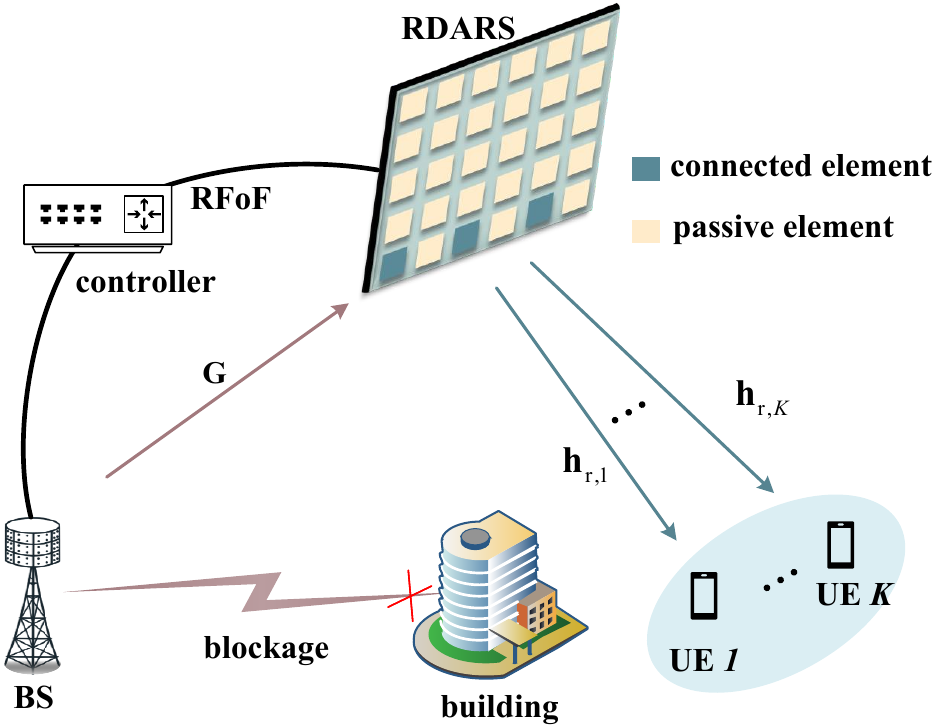}
\caption{An illustration of RDARS-aided communication system, where the connected elements form a uniform sparse array.}
\label{fig: 1_system_architecture}
\vspace{-10pt}
\end{figure}
As shown in Fig. \ref{fig: 1_system_architecture}, we consider a RDARS-aided multi-user downlink system, which consists of a base station (BS) and $K$ single-antenna UEs. The antennas at the BS and the elements of the RDARS are arranged as uniform linear arrays (ULAs), with the inter-element spacing denoted as $d$. The $a$ elements work in the connection mode and the remaining $N -a$ elements work in reflection mode. The elements working in connection modes, i.e., connected elements, are linked with the BS via the radio frequency over fiber (RFoF) \cite{ChengzhiMa_ANewArchi}.
The CEA is formed as a uniform sparse array, where the index set is denoted by
$\mathcal{I} = \{z_m| z_m = m_{0} + m\eta , \eta\in \mathcal{F}, m = 0, 1, \cdots, a-1\}$ with $\mathcal{F} = \{1, 2, \cdots, \left\lfloor {(N-1)}/{(a-1)} \right\rfloor\}$. Let $m_{0}$ denote the index of the reference connected element. 
For the sake of analysis, we set $m_0 = 1$.
The inter-element spacing of the CEA is $\eta d$, with $\eta \in \mathcal{F}$ being the sparsity level \cite{wang2023can, lu2025flexible}.  Let ${\bf A}(\eta) = \operatorname{diag}\{{\mathbf{a}(\eta)}\}$ denote the mode switching matrix, where ${\bf {a}}(\eta) = [a_1, a_2, \cdots, a_N]^{T}$, $a_n =1, n \in \mathcal{I}$, and $a_n =0, n \notin \mathcal{I}$. 
Let $\mathcal{N}_{\setminus N} = \{1, 2, \cdots, N-1\}$, $\mathcal{A} = \{1,2,\cdots, a\}$, and $\mathcal{A}_{\setminus a} =\{1,2,\cdots, a-1\}$, respectively.
Then, the equivalent mode switching matrix $\widetilde{\bf A} (\eta) \in \mathbb{Z}^{N \times a}$ can be constructed by collecting the columns of ${\bf{A}}(\eta)$ including 1. 


The BS-RDARS and RDARS-UE $k$ channels are denoted by $\mathbf{G} \in \mathbb{C}^{N \times N_{\rm t}}$ and $\mathbf{h}_{{\rm r}, k} \in \mathbb{C}^{N \times 1}$, respectively. The beamforming matrices at the BS and RDARS CEA are denoted by ${\bf W} \in \mathbb{C}^{N_{\rm t}\times K}$ and ${\bf F} \in \mathbb{C}^{a\times K}$, respectively. The active beamforming matrix is ${\bf V} = [\mathbf{W}^T, {\bf F}^T]^T$ with ${\bf V} \in \mathbb{C}^{(N_{\rm t}+a) \times K}= [{\bf v}_1, {\bf v}_2, \cdots, {\bf v}_K] $. As a preliminary study on sparsity design for RDARS connected elements, we assume identical effective power for all connected RDARS elements. Let ${\bm \Phi} \in \mathbb{C}^{N\times N}= \operatorname{diag}({{\bm \phi}^H}) $ denote the passive beamforming matrix for RDARS passive elements, where ${\bm \phi} = [e^{j\varphi_1}, e^{j\varphi_2}, \cdots, e^{j\varphi_N}]^{H}$. The symbol vector is denoted by ${\bf s} = [s_1, s_2, \cdots, s_K]^{T}$, where $\mathbb{E}\{s_is_i^H\} = 1$ and $\mathbb{E}\{s_is_j^H\} = 0$ when $i\neq j$. 

Therefore, the received signal for the $k$-th UE is 
${y_k} = {\bf h}_k {\bf v}_k s_{k} + \sum\nolimits_{i = 1,i\ne k}^K {{\mathbf{h}}_{k}{{\mathbf{v}}_i}{s_i}} + n_k$,
where ${\bf h}_k \in \mathbb{C}^{1\times (N_{\rm t} + a)}= [{\mathbf{h}}_{r,k}^H\left( {{\mathbf {I}_N} - {\mathbf{A}(\eta)}} \right){\mathbf{\Phi G}},{\mathbf{h}}_{r,k}^H{\mathbf{\widetilde A}(\eta)}] $ denotes the effective channel of UE $k$, 
and $n_k \sim {\cal C}{\cal N}\left( {0},\sigma_0^2 \right)$ denotes the additive white Gaussian noise (AWGN) for the $k$-th UE with zero mean and covariance $\sigma_0^2$.
The signal-to-noise-plus-interference ratio (SINR) is given by 
$ \gamma_k = \frac{|{\bf h }_k {\bf v}_k |^2}{\sum_{i \neq k}|{\bf h }_k {\bf v}_i|^2 + \sigma_0^2}$. In addition, the sum rate of $K$ UEs is $R({\bf{V}}, {\bf{\Phi}}, \eta) = \sum\nolimits_{k=1}^{K}\log_{2}(1+\gamma_{k})$.
We aim to maximize the sum rate of all the UEs by jointly optimizing active beamforming, passive beamforming, and mode selection matrices. The optimization problem can be formulated as:
\begin{subequations}\label{pro: original sum rate}
 \begin{align}
 \mathop {\max }\limits_{\substack{{{\bf{V}}},{\bf{\Phi}}}, \eta} 
 & \;\; R({\bf{V}}, {\bf{\Phi}}, \eta)
 \\
 \;\textrm{s.t.}\;
 & \operatorname{Tr}({{\bf{V}}}{\bf{V}}^H) \le {P_{\rm{tot}}} ,\label{con: V}\\
 & |\mathbf{\Phi}_{[i,i]}| = 1,  \forall i\in{\mathcal{N}},\label{con: Phi} \\
 & \eta \in \mathcal{F}\label{con: eta},
 \end{align}
\end{subequations}
where $\eqref{con: V}$ and $\eqref{con: Phi}$ denote the total transmit power constraint and the unit-modulus constraint for the active and passive beamforming matrices, respectively. Moreover, the sparsity of the CEA is constrained by $\eqref{con: eta}$. 
It is observed that problem \eqref{pro: original sum rate} is challenging to be directly solved due to the unit-modulus constraint $\eqref{con: Phi}$ and discrete constraint $\eqref{con: eta}$. 
In the following, the special cases with a single and two UEs are first analyzed to gain useful insights. Then, an efficient algorithm is proposed for the general case with an arbitrary number of UEs.
\vspace{-10pt}
\section{Proposed Algorithm}
\subsection{Special Cases With the Single-UE and Two-UE}
Let ${\bf b}(N, u) = [1, e^{\frac{2\pi}{\lambda}  u \cdot d}, \cdots, e^{j\frac{2\pi}{\lambda}  u \cdot(N-1)d}]^{T}$ denote the steering function, where $u$ denotes spatial frequency, and $\lambda$ denotes the wavelength. 
The sparse steering vector $\widetilde{{\bf b}}_{\eta}(N, u)$ inherits the entries of ${\bf b}(N, u)$ solely at the positions where the entry of ${\bf a}(\eta)$ is equal to 1, with all other entries set to zero. 
For the basic line-of-sight (LoS) propagation, the channel of BS-RDARS link is 
$ {\mathbf{G}} = {\kappa _{\mathrm{br}}}{\mathbf{b}}( {N,u_{\mathrm{br}}^{\rm{AoA}}} ){{\mathbf{b}^H}}( {{N_{\rm t}},u_{\mathrm{br}}^{\rm{AoD}}} ) \in {\mathbb{C}^{N \times {N_{\rm t}}}}$ with $u_{\mathrm{br}}^{\rm{AoA}} = \cos (\theta^{\rm AoA}_{\rm br})$ and  $u_{\mathrm{br}}^{\rm{AoD}} = \cos (\theta^{\rm AoD}_{\rm br})$, where ${\kappa _{\mathrm{br}}}$, $\theta^{\rm AoA}_{\rm br}$  and $\theta^{\rm AoD}_{\rm br}$ denote the path loss coefficient, angle-of-arrival (AoA), and angle-of-departure (AoD), respectively. The channel of RDARS-UE $k$ link is $ {{\mathbf{h}}_{{\rm r}, k}} = {\kappa _{\mathrm{ru}, k}}{\mathbf{b}}( {N,u_{\mathrm{ru},k}^{\rm AoD}} ) \in {\mathbb{C}^{N \times 1}}$ with $u_{\mathrm{ru}, k}^{\rm AoD} = \cos (\theta^{\rm AoD}_{{\rm ru}, k})$, where ${\kappa _{\mathrm{ru}, k}}$ and $\theta^{\rm AoD}_{{\rm ru},k}$ denote the path loss coefficient and the AoD from the RDARS to UE $k$, respectively.
\subsubsection{Single UE}
The single-UE case is first analyzed as a baseline to draw useful insights for the proposed design.
The received signal of the single UE is 
$y = {\mathbf{h}}_{\rm r}^H\left( {{\mathbf{I}} - {\mathbf{A}}} \right){\mathbf{\Phi Gw}}s + {\mathbf{h}}_{\rm r}^H{\mathbf{\widetilde Af}}s + {n} \nonumber= \mathbf{h} \mathbf{v}s + {n}$.
Problem \eqref{pro: original sum rate} is reduced to 
\begin{align}
 \label{pro: SNR maximization}
 \mathop {\max }\limits_{\substack{{{\bf{w}}},{\bf{f}},{\bf{\Phi }}, \eta} } 
  \; \gamma
  \;\;\; \textrm{s.t.}\;
  \operatorname{Tr}({{\bf{v}}}{\bf{v}}^H) \le {P_{\rm{tot}}},
  \eqref{con: Phi}, \eqref{con: eta}. 
  \vspace{-5pt}
\end{align}

Then, the optimal phase shift and beamforming vectors are ${\varphi}_n^{{\rm opt}} = \arg ({[{\mathbf{b}}( {N,u_{\mathrm{ru}}^{\rm AoD}})]_n}) - \arg ({[{\mathbf{b}}( {N,u_{\mathrm{br}}^{\rm AoA}} )]_n})$, ${{\mathbf{w}}^{\rm opt}} = \frac{\sqrt {{P_{\rm b}^{\rm opt}}} {{\mathbf{b}}\left( {{N_{\rm t}},u_{\mathrm{br}}^{\rm AoD}} \right)}}{{||{\mathbf{b}}\left( {{N_{\rm t}},u_{\mathrm{br}}^{\rm AoD}} \right)|{|_2}}}$, and ${{\mathbf{f}}^{\rm opt}} =  \widetilde {\mathbf{ A}}^{T}(\eta)\frac{{{\sqrt {{P_{\rm r}^{\rm opt}}}\mathbf{b}}\left( {N,u_{\mathrm{ru}}^{\rm AoD}} \right)}}{{||{\mathbf{b}}\left( {N,u_{\mathrm{ru}}^{\rm AoD}} \right)|{|_2}}}$. With the total transmit power constraint, the optimal power allocation can be obtained, given by  
$P_{\rm b}^{\rm opt} \!=\! \frac{{{\kappa _{\mathrm{br}}}(N - a)\sqrt {{N_{\rm t}P_{\rm tot}}} }}{{\sqrt {\kappa _{\mathrm{br}}^2{{(N - a)}^2}{N_{\rm t}} + a} }}$ and $P_{\rm r}^{\rm opt} \!=\!  \frac{{\sqrt{P_{\rm tot}a}}}{{\sqrt {\kappa_{\mathrm{br}}^2{{(N - a)}^2}{N_{\rm t}} + a}}}$.
Therefore, the maximum SNR is ${\gamma _{\max }} = \frac{{\kappa _{\mathrm{ru}}^2{P_{\rm tot}}(\kappa _{\mathrm{br}}^2{{(N - a)}^2}{N_{\rm t}} + a)}}{{{\sigma_0^2}}}$,
which only depends on the numbers of total RDARS elements and connected elements, as well as the total transmit power, while irrespective of the sparsity of the CEA. As a result, any CEA placement can achieve the maximum achievable rate in the single-UE case.
\subsubsection{Two UEs}
Next, based on the single-UE analysis, we extend the discussion to the two-UE case.
For the case of two UEs, the maximum-ratio transmission (MRT), zero-forcing (ZF) and minimum MSE (MMSE) beamforming schemes are considered.
Let $\varepsilon_{k,{k'}} = \frac{\left|{\mathbf{h}}_k {\mathbf{h}}^{H}_{k'}\right|^2}{\left|\left|{\bf{h}}_k \right| \right|^2\left|\left|{\mathbf{h}}_{k'} \right| \right|^2 }$ denote the CSCC, where $k,k' = 1,2, k\neq k'$. Let $p_{k}$ denote the UE transmit power for the $k$-th UE, and $\beta_{k} = ||\mathbf{h}_{k}|| = \sqrt{\xi _k^2{N_{\rm t}} + \kappa _{{\rm \mathrm{ru}},k}^2a}$ with ${\xi_k}\!=\!\kappa _{\mathrm{br}}\kappa_{\mathrm{ru},k}(\sum\limits_{n = 1}^N \!\!e^{j( {{\varphi _n} \!+ \!\frac{2\pi d}{\lambda} ( {n  -  1} )\Delta {u_k}} )}\!\! -\!\!\! \sum\limits_{m = 0}^{a - 1} {{e^{j( {{\varphi _{{n_m}}}\! \!+\!\! \frac{2\pi d}{\lambda} ( {{n_m}  -  1} )\Delta {u_k}} )}}})$. Therefore, the SINRs under the three beamforming schemes are given by
$\gamma _k^{\rm{MRT}}  = \frac{{p_k}\beta _k^2}{{\sigma_0 ^2}} ( {1 - \frac{{{p_{{k'}}}\beta _k^2{\varepsilon _{k,{k'}}}/{\sigma_0 ^2}}}{{1 + {p_{{k'}}}\beta _k^2{\varepsilon _{k,{k'}}}/{\sigma_0 ^2}}}})$, $\gamma _k^{\rm{ZF}} = \frac{{{p_k}\beta _k^2}}{{{\sigma_0 ^2}}}( {1 - {\varepsilon _{k,{k'}}}} )$, and $\gamma _k^{{\rm{MMSE}}}\!\! =\!\! \frac{{{p_k}\beta _k^2}}{{{\sigma_0 ^2}}}( {1 \!\!-\!\! \frac{{{p_{{k'}}}\beta _{{k'}}^2/{\sigma_0 ^2}}}{{1 + {p_{{k'}}}\beta _{{k'}}^2/{\sigma_0 ^2}}}{\varepsilon _{k,{k'}}}} )$ \cite{haiquan_survey}.

A closer look at the three SINR expressions shows that a larger SINR can be achieved by decreasing $\varepsilon_{k,{k'}}$.
Moreover, the terms $\frac{{{p_{{k'}}}\beta _k^2{\varepsilon _{k,{k'}}}/{\sigma_0 ^2}}}{{1 + {p_{{k'}}}\beta _k^2{\varepsilon _{k,{k'}}}/{\sigma_0 ^2}}}$, ${\varepsilon _{k,{k'}}}$, and $\frac{{{p_{{k'}}}\beta _{{k'}}^2/{\sigma_0 ^2}}}{{1 + {p_{{k'}}}\beta _{{k'}}^2/{\sigma_0 ^2}}}{\varepsilon _{k,{k'}}}$ account for the SNR loss factors for UE $k$ due to applying the MRT, ZF, and MMSE beamforming schemes, respectively.

In this case, we aim to minimize $\varepsilon_{k,{k'}}$, and the problem is 
\begin{align}\label{pro: channel correlation minimization}
 \mathop {\min }\limits_{\substack{\eta, {\bm \Phi}} } 
 \;\; \varepsilon_{k,{k'}} 
 \;\;\;\textrm{s.t.}\;  \eqref{con: Phi},  \eqref{con: eta}.
\end{align}
Specifically, we have $\mathbf{h}_k \mathbf{h}_{k'}^{H}  = {\xi_k}\xi_{k'}^*{N_{\rm t}} + {\kappa _{\mathrm{\mathrm{ru}},k}}{\kappa_{\mathrm{\mathrm{ru}},{k'}}}{{{\bar S}}_{\eta} }(\Delta u)$, with
${\Delta u} \!=\! u_{{\rm {\mathrm{ru}}}, 2}^{\rm AoD}  -  u_{{\rm {\mathrm{ru}}}, 1}^{\rm AoD}$, 
$\Delta {u_k} = u_{\mathrm{br}}^{\rm AoA} - u_{\mathrm{ru},k}^{\rm AoD}$, and
${\bar{S}_\eta }(\Delta u) = {S_\eta }(\Delta u) {\psi _{a,\eta }}( \Delta  u)$ with ${S_\eta }(\Delta u) ={{\sin (\frac{{a\pi d \eta \Delta u}}{\lambda})}}/{{\sin (\frac{{\pi d\eta \Delta u}}{{\lambda} })}} $ and ${\psi _{a,\eta }}( \Delta  u) =e^{j\frac{2\pi d}{\lambda}({{m_0} - 1 +{\frac{{a - 1}}{2}}\eta})\Delta u} $. We denote $\xi_{k'}^*$ as the conjugate of $\xi_{k'}$.
Let ${D_N}(\Delta {u_k};{\mathbf{\Phi }})  =  \sum\nolimits_{n = 1}^N {{e^{j ( {{\varphi _n} + \frac{2\pi d}{\lambda} \left( {n - 1} \right)\Delta {u_k}} )}}}$ and ${\widetilde{S}_\eta }(\Delta u_k; {\bm {\Phi}} ) = \sum\nolimits_{m = 0}^{a - 1} {{e^{j\left( {{\varphi_{{m_0} + m\eta }} + \frac{2\pi d}{\lambda} \left( {{m_0} + m\eta  - 1} \right)\Delta u_k } \right)}}}$. 
The CSCC of two UEs is given by ${\varepsilon _{1,2}}( \eta; {\mathbf{\Phi }}) = {{{\left| { D_1 D^*_2  + S_{1,2}   } \right|}^2}}    /{({{({|D_1|^2 + \kappa _{\mathrm{ru},1}^2a})(  |D_2|^2  + \kappa _{\mathrm{ru},2}^2a})})}$, where $D_{k} = \sqrt{N_{\rm t}}{\kappa_{\mathrm{br}}}{\kappa _{\mathrm{ru},k}}({D_N}(\Delta {u_k};{\mathbf{\Phi}}) -{\widetilde{S}_\eta }(\Delta u_k;{\mathbf{\Phi }} )) $, $S_{1,2} = {\kappa _{\mathrm{ru},1}}{\kappa_{\mathrm{ru},2}} {S_\eta }\left( \Delta u \right){e^{  j{\psi _{a,\eta }}( \Delta  u)}}$.
It can be seen that the sparsity level is determined by system parameters such as the numbers of BS antennas and RDARS elements, path loss, passive beamforming, and the phase difference between UEs. 

Note that when the passive beamforming points towards the arbitrary reference direction ${u_{\rm ref}}$, with ${\varphi _n} = \frac{2\pi d}{\lambda} (n - 1)({u_{\rm ref}} - u_{\rm \mathrm{br}}^{\rm AoA})$, the full-array Dirichlet beampattern is
${\bar{D}_N}(\Delta {\widetilde{u}_k}) \!\! =\!\! \sum\limits_{n = 1}^N \!{{e^{j( {\frac{2\pi d}{\lambda} ( {n - 1} ) \Delta {\widetilde{u}_k} } )}}} = \frac{{\sin ({{N\pi d \Delta {\widetilde u_k}}})}}{{\sin ({{\pi d \Delta {\widetilde u_k}}})}}{e^{j\widetilde{\psi}_k}}$, with $\widetilde{\psi}_k = \frac{\pi d}{\lambda} ( {{N\! - \!1}})\Delta {\widetilde{u}_k}$ and ${\Delta {\widetilde{u}_k}} = {u_{\rm ref}}\! -\! u_{{\rm {\mathrm{ru}}}, k}^{\rm AoD}$.

In the following, the three cases are respectively considered.

\textbf{Case 1:} When {${\varphi _n} \!=\!\! \frac{2\pi d}{\lambda} \! (n \!-\! 1) ( \! {u_{\rm ref}} \!-\! u_{\rm \mathrm{br}}^{\rm AoA} \!)$}, the impact of the sparsity on CSCC is analyzed under two following subcases.

\textbf{Subcase 1:} 
When ${(N + a)^2N_{\rm t}}/{a} \ll {1}/{\kappa_{\mathrm{br}}^{2}}$, we have 
$ \frac{|\widetilde{D}_{k} \widetilde{D}^*_{k'} |}{{| {{\kappa _{\mathrm{ru},k}}{\kappa _{\mathrm{ru},k'}}{{\bar S}_\eta }(\Delta u)}|}} \ll 1$, 
where $\widetilde{D}_k \!\!=\!\! \sqrt{N_{\rm t}}\kappa _{\mathrm{br}}{\kappa _{\mathrm{ru},k}}{\widetilde{D}_N}(\Delta {\widetilde{u}_k})$ with ${\widetilde{D}_N}(\Delta {\widetilde{u}_k})\!\! = \!\!{{{\bar{D}}_N}(\Delta {\widetilde{u}_k}) - {{\bar S}_\eta }(\Delta \widetilde{u}_k)}$. In this case, the RDARS-UE link dominates the performance, and 
$ {\varepsilon _{1,2}} ( \eta )\!\! \approx \!\!\left| {\kappa _{\mathrm{ru},1}}{\kappa _{\mathrm{ru},2}}{S_\eta }\left( \Delta u \right){e^{  j{\psi _{a,\eta }}\left( \Delta  u\right)}} \right|^2/{{( {\kappa _{\mathrm{ru},1}^2} {\kappa _{\mathrm{ru},2}^2a^2} )}}\!\! =\!\! \frac{{{{| {{S_\eta }( \Delta u )} |}^2}}}{a^2} $.
As such, the optimal $\eta$ is given by $\eta^{\rm opt} \in \mathcal{R}$, where $\mathcal{R}\!\triangleq\!\!\{\operatorname{round}({\frac{{q\lambda}}{{ad\Delta u }}})|q \in \mathcal{A}_{\setminus a}\} \cap \mathcal{F}$. Thus, the maximum SINR can be achieved when the CEA is uniformly sparse with $\eta \in \mathcal{R}$.

\textbf{Subcase 2:} When ${{{( {N + a})}^2}{N_{\rm t}}}/{a} \gg {1}/{{\kappa _{\mathrm{br}}^2}}$, we have
${| {N_{\rm t}}\kappa _{\mathrm{br}}^2 {\widetilde{D}_N}(\Delta {\widetilde{u}_1})  {{\widetilde{D}^*_N}(\Delta {\widetilde{u}_2})}|}/{{|{{\bar S}_\eta }(\Delta u)|}}\gg 1$. 
In this case, the reflection link dominates the performance. In this case, we have 
${\varepsilon _{1,2}} ( \eta  ) = 1$. 
As a result, the DoF of reflection link is insufficient to spatially distinguish two UEs.
Therefore, the optimal sparsity is $\eta^{\rm opt} \in \mathcal{F}$, which indicates that the maximum SINR can be achieved when the sparsity level is arbitrarily selected from $\mathcal{F}$.

\textbf{Case 2:} In this case, we set the spatial frequency of RDARS passive elements $u_{\rm ref}$ as the average of the two UEs' spatial frequency $u^{\rm AoD}_{{\rm {\mathrm{ru}}},1}$ and $u^{\rm AoD}_{{\rm {\mathrm{ru}}},2}$ for the sake of fairness, where
$\varphi^{\rm opt}_{k,n} = \frac{2\pi d}{\lambda} (n-1) (u^{\rm AoD}_{{\rm {\mathrm{ru}}},k} - u^{\rm AoA}_{{\rm {\mathrm{br}}}})$ for $k=1,2$.
As such, we have $\varphi^{\rm ref}_{n} = ({\varphi^{\rm opt}_{1,n} + \varphi^{\rm opt}_{2,n}})/{2} = \frac{2\pi d}{\lambda} (n-1) (\frac{u^{\rm AoD}_{{\rm {\mathrm{ru}}},1}+u^{\rm AoD}_{{\rm {\mathrm{ru}}},2}}{2} - u^{\rm AoA}_{{\rm {\mathrm{br}}}})$, where $u_{\rm ref} = ({u^{\rm AoD}_{{\rm {\mathrm{ru}}},2}+u^{\rm AoD}_{{\rm {\mathrm{ru}}},1}})/{2}$. Let $D_{N}(\Delta u) = \frac{\sin(\pi dN\Delta u/\lambda)}{\sin(\pi d\Delta u /\lambda)}$. In this case, the CSCC is ${\bar \varepsilon _{1,2}} (\eta)\! = {{{ {{\kappa ^2_{\mathrm{ru},1}}{\kappa^2_{\mathrm{ru},2}}}}{|\widetilde{X} + {S_\eta }(\Delta u){e^{j{\psi _{a,\eta }}\left( {\Delta u} \right)}} |^2}}} /{({\bar{X}_{1} \bar{X}_{2} })}$
where $X({\theta _\eta }(\Delta u)) = \! {D_N}( {{{\Delta u}}/{2}}){e^{j{\psi _N}( {\frac{{\Delta u}}{2}} )}} - S_\eta ( {{{\Delta u}}/{2}} ) {e^{j{\psi _{a,\eta }}( {\frac{{\Delta u}}{2}} )}}$ with ${\psi _N}( \frac{\Delta  u}{2}) = \frac{\pi d}{\lambda}(N-1)\frac{\Delta  u}{2}$, $\widetilde{X} = {N_{\rm t}}\kappa _{\mathrm{br}}^2| {X({\theta _\eta }(\Delta u))}|^2 {e^{j2\operatorname{arg}({X({\theta _\eta }(\Delta u))})}}$, and $\bar{X}_{k}  = \kappa _{\mathrm{ru},k}^2a +  |{{\kappa _{\mathrm{br}}}{\kappa _{\mathrm{ru},k}}}|^2{{\left| {X({\theta _\eta }(\Delta u))} \right|}^2}{N_{\rm t}}$. It is observed from ${\bar \varepsilon _{1,2}}\left( \eta  \right)$ that the SINR hinges on the sparsity of the CEA, where the sparsity alters both the amplitude and phase of ${\bar \varepsilon _{1,2}}\left( \eta  \right)$. Therefore, it is difficult to derive the optimal $\eta$ in closed-form, and the optimal $\eta$ can be obtained numerically, i.e., $\eta^{\rm opt} = \arg\min_{\eta\in \mathcal{F}}{\bar \varepsilon _{1,2}}\left( \eta  \right)$.

\textbf{Case 3:} 
When $\Delta u = 0$, the CSCC is given by ${\varepsilon _{1,2}}\left( \eta  \right) = 1$. This indicates that the channel for two UEs cannot be distinguished in the spatial domain, thus resulting in a severe inter-user interference (IUI) issue. 
Therefore, the sparsity level can be arbitrarily selected from $\mathcal{F}$.  

Based on the above results, we have Proposition \ref{propo 1}.
\begin{Proposition} \label{propo 1}
Under the assumption {${\varphi _n} =\frac{2\pi d}{\lambda} (n - 1)({u_{\rm ref}} - u_{\rm \mathrm{br}}^{\rm AoA})$}, the optimal sparsity is obtained as 
\begin{align} \label{equ: optimal eta}
\eta^{\rm opt} \in 
\begin{cases}
\mathcal{F}, \text{if } {(N + a)^2N_{\rm t}}/{a} \gg {1}/{\kappa_{\mathrm{br}}^{2}}\ \text{or}\  \Delta u = 0,\\
\mathcal{R},  \text{if } {(N + a)^2N_{\rm t}}/{a} \ll {1}/{\kappa_{\mathrm{br}}^{2}}, \\
\arg\min_{\eta\in \mathcal{F}}{\bar \varepsilon _{1,2}}\left( \eta  \right), \text{if } \varphi_n = \varphi^{\rm ref}_{n}, n \in \mathcal{N}.
\end{cases}
\vspace{-5pt}
\end{align}
\end{Proposition}

Note that sparsity design can mitigate the IUI by minimizing the CSCC.
With \eqref{equ: optimal eta}, we have the minimum CSCC, and the corresponding SINRs are close to the SNR without IUI. Thus, the maximum sum rate of two UEs can be achieved by setting the sparsity level according to Proposition \ref{propo 1}. 
In particular, compared to the compact array with half-wavelength element spacing, the uniform sparse element array can achieve a larger array aperture, thus resulting in a narrower beamwidth for distinguishing the UEs spatially.
\vspace{-10pt}
\subsection{Arbitrary Number of UEs}
In this subsection, the general case with an arbitrary number of UEs is considered, where a WA algorithm is proposed to tackle the non-convex optimization problem \eqref{pro: original sum rate}. By introducing the auxiliary vectors $\bm{\zeta}=[\zeta_1, \cdots, \zeta_K]^{T}$ and ${\bf{u}} = [\mu_1, \cdots, \mu_K]^{T}$,
problem \eqref{pro: original sum rate} can be transformed into a WMMSE problem: 
\begin{align}\label{pro: sum rate MMSE}
 \mathop {\min }\limits_{\substack{{{\bf{V}}},{\bf{\Phi }},\eta, \bf{u}, \bm{\zeta}}}\;
 \sum\nolimits_{k = 1}^K {({\zeta _k}{e_k} - \log {\zeta _k})}
\;\;\textrm{s.t.}\;
 \eqref{con: V}, \eqref{con: Phi}, \eqref{con: eta}, 
\end{align}
where $e_k$ denotes the MSE for UE $k$, given by ${e_k} = 1 \!\!-\!\! \mu_k^H{{\bf{h}}_k}{{\bf{v}}_{k}}\!\! -\!\!{\bf{v}}_{k}^H{\bf{h}}_k^H{u_k}\!\! + \!\! \mu_k^H{{\bf{h}}_k}\sum\nolimits_{m = 1}^K {{{\bf{v}}_{m}}{\bf{v}}_{m}^H} {\bf{h}}_k^H{\mu_k}\!\! + \!\!\mu_k^H{ \mu_k}{\sigma_{0}^{2}}$ \cite{Qi_WMMSE_Algo}.

\subsubsection{Active Beamforming Optimization}
Given $\{{\bm \Phi}, \eta\}$, the sub-problem with respect to ${\bf {V},{\bf u}, {\bm \zeta}}$ can be expressed as 
\begin{align}\label{pro: sum rate MMSE fixed Phi, A}
\mathop {\min }\limits_{\substack{{{\bf{V}}, {\bf u}, {\bm \zeta}}}}\;
  \sum\nolimits_{k = 1}^K {({\zeta _k}{e_k} - \log {\zeta _k})} \;\;\textrm{s.t.}\; \eqref{con: V}.
 \end{align}
By taking the derivative of the objective function in \eqref{pro: sum rate MMSE fixed Phi, A} with respect to each variable, we can obtain the solutions in closed-form, given by
$\mu_k\!\!= \!\!{({{\bf{h}}_k}\sum\nolimits_{m = 1}^K {{{\bf{v}}_{m}}{\bf{v}}_{m}^H} {\bf{h}}_k^H + \frac{{\sigma_{0}^{2}}}{P_{\rm{tot}}}\sum\nolimits_{m = 1}^K {{\bf{v}}_{m}^H} {{\bf{v}}_{m}})^{-1}}{{\bf{h}}_k}{{\bf{v}}_{k}}$, $\zeta _k = {e_{k}^{-1}}$, and 
${\bf{v}}_{k}= {\mu_k}{\zeta _k}(\!\sum\nolimits_{m = 1}^K {\mu_m^H{\mu_m}{\zeta _m}(\rho{{\bf{I}}}\!\! + \!\!{\bf{h}}_m^H{{\bf{h}}_m}))^{-1}{\bf{h}}_k^H}$,
where $\rho \ge 0$ is the Lagrange multiplier with respect to  the power constraint and can be obtained by the bisection search such that $\operatorname{Tr}({\bf V}{\bf V}^H) = P_{\rm tot}$. In addition, the active beamforming matrix is initialized with the ZF method.

\subsubsection{Passive Beamforming Optimization}
Given $\{{\bf V}, \eta\}$, the sub-problem for optimizing the passive beamforming is
\begin{align}
 \mathop {\min }\limits_{\bm{\phi}} \quad {{\bm{\phi }}^H}\bf{C}{\bm{\phi }} + {{\bm{\beta }}^H}{\bm{\phi }} + {{\bm{\phi }}^H}{\bm{\beta }}
 \;\;\;\;\;\;\;\textrm{s.t.}\;\; \eqref{con: Phi},\label{pro: passive BF}
 \end{align}
where ${\bf{C}} = \sum\nolimits_{k = 1}^{K} {\zeta _k}\mu_k^H{\mu_k}\bar{\bf A}{{\bf{H}}_{{\rm{r}},k}}\sum\nolimits_{m = 1}^K {{\bf{w}}_{m}}{\bf{w}}_{m}^H {{\bf{H}}^{H}_{{\rm{r}},k}} ({\bf{I}}_{N}- {\bf{A}})$ and ${\bm{\beta }} = \sum\nolimits_{k = 1}^K { {\zeta _k}\mu_k^H{\mu_k}\bar{\bf A}{{\bf{H}}_{{\rm{r}},k}}\sum\nolimits_{m = 1}^K {{{\bf{w}}_{m}}{{\bf{f}}^{H}_{m}}{{{\bf{\widetilde A}}}^H}} } {\bf{h}}_{{\rm{r}},k}- {\zeta _k}\mu_k^H{\bar{\bf A}}{{\bf{H}}_{{\rm{r}},k}}{{\bf{w}}_{k}}$ with ${\bar{\bf A}} = {\bf{I}}_{N}- {\bf{A}}$.
Let $\mathbf{p}= [\bm{\phi}, q]^T$, where $q$ is an auxiliary variable. Thus, problem \eqref{pro: passive BF} is equivalent to
\begin{align}\label{pro: rankone}
 \mathop {\max}\limits_{\bf{p}} 
 \;\;{{\bf{p}}^H}{\bf{Dp}}
 \;\;\textrm{s.t.}\;\;|p_n| = 1, n= 1,\cdots, N+1,
\end{align}
where ${\bf{D}} = \left[ { - {\bf{C}}}, { - {\bm{\beta}}};{- {{\bm{\beta }}^H}}, 0 \right]$, and $p_n$ denotes the $n$-th element of ${\bf{p}}$.
It is observed that the optimization problem \eqref{pro: rankone} can be solved by the power iteration algorithm. Specifically, the value of $p$ in the $q$-th iteration is 
$  {\bf{p}}^{(q+1)}= e^{j\mathrm{arg}(({\bf{D}}+\nu {\bf{I}}_{N+1}){\bf{p}}^{(q)})}$,
where $\nu {\bf{I}}_{N+1}$ is introduced to ensure that ${\bf{D}}+\nu {\bf{I}}_{N+1}$ is a positive definite matrix. The passive beamforming vector can be updated based on $\bm{\phi} = e^{j \arg(\frac{\mathbf{p}{[1:N]}}{p_{N+1}} )}$ until the convergence is achieved.
\subsubsection{Sparsity Optimization}
 Given $\{{\bf V}, {\bm \Phi}\}$, the sub-problem with respect to $\eta$ is 
 \begin{align}\label{pro: sum rate MMSE fixed Phi, V}
\mathop {\min }\limits_{\substack{{\eta}}}\;
 & \sum\nolimits_{k = 1}^K {({\zeta _k}{e_k} - \log {\zeta _k})} \;\;\;\;\;\;\;\textrm{s.t.}\;\; \eqref{con: eta}.
 \end{align}
Since the sparsity $\eta$ impacts the objective function in a  intricate manner, which cannot be directly optimized. Fortunately, the optimal $\eta$ can be obtained by searching over the set $\mathcal{F}$, given by
$\eta^{\rm opt} = \arg \operatorname{max}_{\eta} R({\bf{V}}, {\bf{\Phi}}, \eta)$.

Thus, by updating the variables of active beamforming and passive beamforming in an iterative manner and obtaining the optimal sparsity via exhaustive search, problem \eqref{pro: original sum rate} can be solved until the fractional increase of the objective value is below a threshold, i.e., $10^{-4}$.

\subsubsection{Computational Complexity Analysis}
For the active beamforming optimization, the computational complexity is $\mathcal{O}\!\left(K(N_{\rm t}+a)^{3}\right)$. For the passive beamforming optimization, 
the computational complexity mainly lies in calculating $\mathbf{C}$, yielding a complexity of $\mathcal{O}(K^{2}N^{2})$. In addition, the complexity of the power iteration is $\mathcal{O}(I_{\rm in} N^{2})$, where $I_{\rm in}$ denotes the number of inner iterations required for convergence.
Regarding the mode selection, the complexities associated with optimizing $\mathbf{A}$ and $\widetilde{\mathbf{A}}$ are  $\mathcal{O}((2N)^{3})$ and $\mathcal{O}(N^{3})$, respectively \cite{zhang_RDARS,ji2025model}. 
The computational complexity for sparsity optimization is 
$\mathcal{O}(\lfloor \frac{N-1}{a-1} \rfloor)$.
Thus, the overall computational complexity is  
$\mathcal{O}(I_{\rm out}\lfloor \frac{N-1}{a-1} \rfloor (K(N_{\rm t}+a)^{3} + K^{2}N^{2} + I_{\rm in}N^{2}))$, where $I_{\rm out}$ denotes the number of outer iterations. In addition, $I'_{\rm out}$ and $I''_{\rm out}$ denote the numbers of outer iterations for PWM and MM algorithms, respectively.
A detailed comparison of the complexities is provided in Table~\ref{tab 1: complexity comprasion}. 
For example, when $N_{\rm t} = 16$, $N = 512$, $a=4$, $K=4$, $I_{\rm out} = I'_{\rm out} =I''_{\rm out}=10$, $I_{\rm in}=3$, the computational complexity of WA is reduced by 62.72\% and 29.75\% relative to those of the MM and PWM, respectively.
\begin{table}
\scriptsize
\caption{Complexity and Runtime Analysis*}
\vspace{-5pt}
\centering
\begin{tabular}{ccc}
\toprule
\textbf{Algorithm}&\textbf{Computational Complexity}&\textbf{Runtime (s)} \\
\midrule
MM \cite{zhang_RDARS}&$\mathcal{O}(I''_{\rm{out}}(K(N_{\rm{t}} + a)^{3.5} + K^2 N^2 + 17 N^3) )$&17.416 \\
PWM \cite{ji2025model}&$\mathcal{O}(I'_{\rm out}(K(N_{\rm{t}} \!+\! a)^3 \!+\! (I_{\rm{in}} \!+\! K^2 )N^{2} \!+\! 9 N^3) )$&10.3234\\
WA &$\mathcal{O}(I_{\rm{out}}\lfloor\frac{N-1}{a-1}\rfloor (K(N_{\rm{t}} \!+\! a)^3 \!\!+\!\! (I_{\rm{in}} \!+\! K^2 )N^{2}))$&7.6657\\ 
\bottomrule
\end{tabular}
\label{tab 1: complexity comprasion}
\begin{tablenotes}
\footnotesize
\item *The runtime results are averaged over 500 Monte Carlo trails, with all simulations conducted on a PC equipped with an 11th Gen Intel Core i7-11700 CPU.
\end{tablenotes}
\vspace{-15pt}
\end{table}

\vspace{-10pt}
\section{Numerical Results}
In this section, numerical results are provided to verify the effectiveness of the sparsity design. The BS and RDARS are located at (0,0,15) m and (50,30,15) m, respectively. The UEs are randomly distributed within a circle, where its center and radius are (100, 0, 1.5) m and 20 m, respectively. We set $N_{\rm{t}} =32$, $K = 20$, $N = 128$, $P_{\rm{tot}} = 30$ dBm, $f_{0} =28$ GHz, $a = 20$, $c_0 = 61.4$ dB, $d = \frac{\lambda}{2}$, and $\sigma^2_k = -91.4$ dBm, where $c_0$ denotes the path loss at the reference distance 1 m. The performance is evaluated over 1,000 Monte Carlo trials. For comparison, the scheme of compact array, i.e., $\eta =1$, and random sparsity scheme are considered.



\begin{figure*}[ht]
\vspace{-20pt}
        \centering
	\begin{minipage}[t]{\linewidth}
        \subfloat[$\eta^{\rm opt} \in \mathcal{R}$.]{\label{fig: 2_solution2}\includegraphics[width=0.33\textwidth]{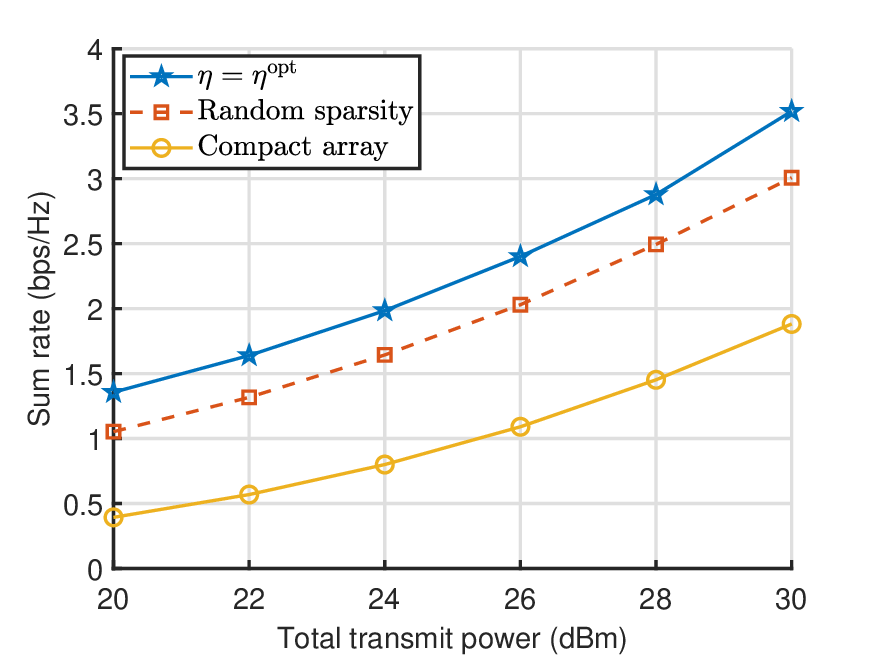}}
        \hspace{-0.13in}
        \subfloat[$\eta^{\rm opt}$ when $\varphi_n = \varphi^{\rm ref}_n$.]{\label{fig: 2_solution4}\includegraphics[width=0.33\textwidth]{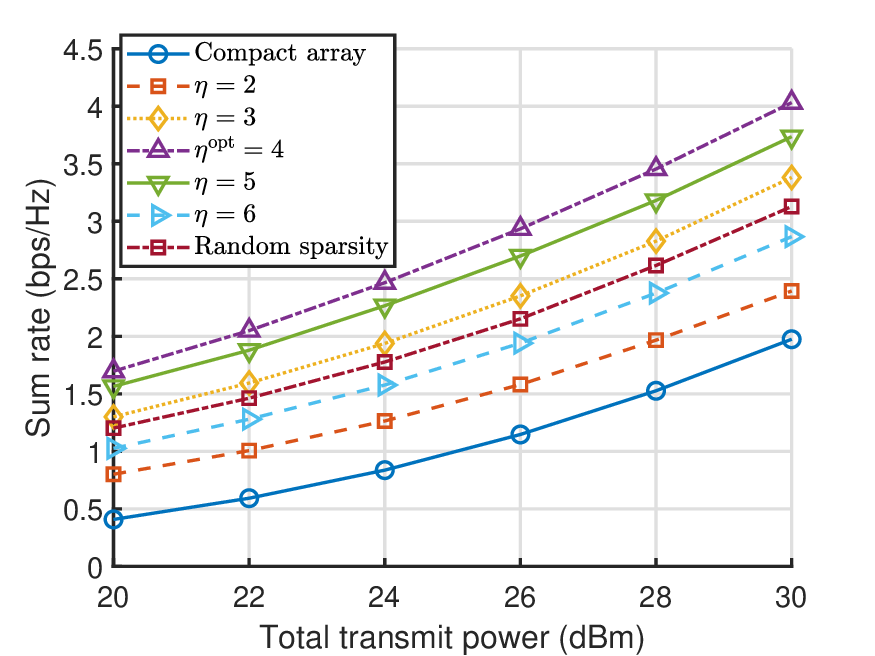}}
        \subfloat[$\eta^{\rm opt} \in \mathcal{F}$.]
        {\label{fig: 2_solution1}\includegraphics[width=0.33\textwidth]{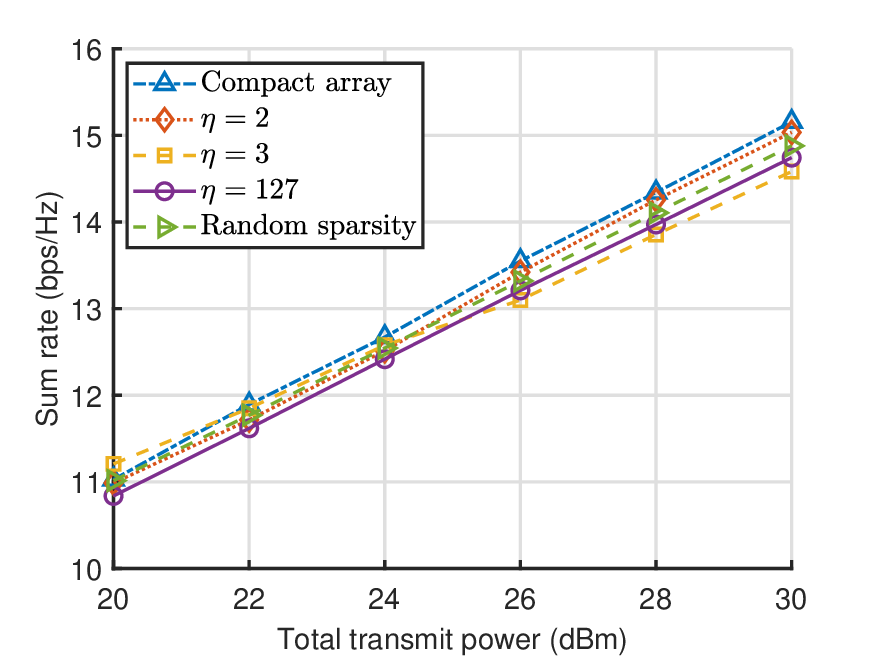}}
        \hspace{-0.13in}
         \caption{Sum rate versus the total transmit power under different cases for two UEs.}
		\label{fig2: two UEs}
	\end{minipage}
\vspace{-10pt}
\end{figure*}


Fig. \ref{fig2: two UEs} shows the sum rate versus the total transmit power under different cases with $K = 2$. Fig. \eqref{fig: 2_solution2} shows the performance comparison for the schemes with $\eta^{\rm opt} \in \mathcal{R}$, random sparsity, and compact array. It is observed that 86.98\% and 17\% rate improvements are achieved at $P_{\rm tot} = 30$ dBm with the proposed solution, respectively. This is because the sparse array yields a higher spatial DoF to reduce the CSCC. 
Fig. \eqref{fig: 2_solution4} shows the sum rate with different sparsity levels under Case 2. 
The sparse CEA outperforms the compact array with $\eta = 1$, thanks to the enlarged physical array aperture. It is also observed that the sum rate initially increases and subsequently decreases as the sparsity increases. This is because the grating lobe issue becomes more severe as $\eta$ increases, which exacerbates the IUI. The above results verify the necessity of optimizing the sparsity level for the RDARS system.
In Fig. \eqref{fig: 2_solution1}, 
the fluctuations in the sum rate caused by the sparsity level are negligible, thereby demonstrating the effectiveness of theoretical analysis under Subcase 2.

\begin{figure}[t] 
 \centering
\includegraphics[width=0.35\textwidth]{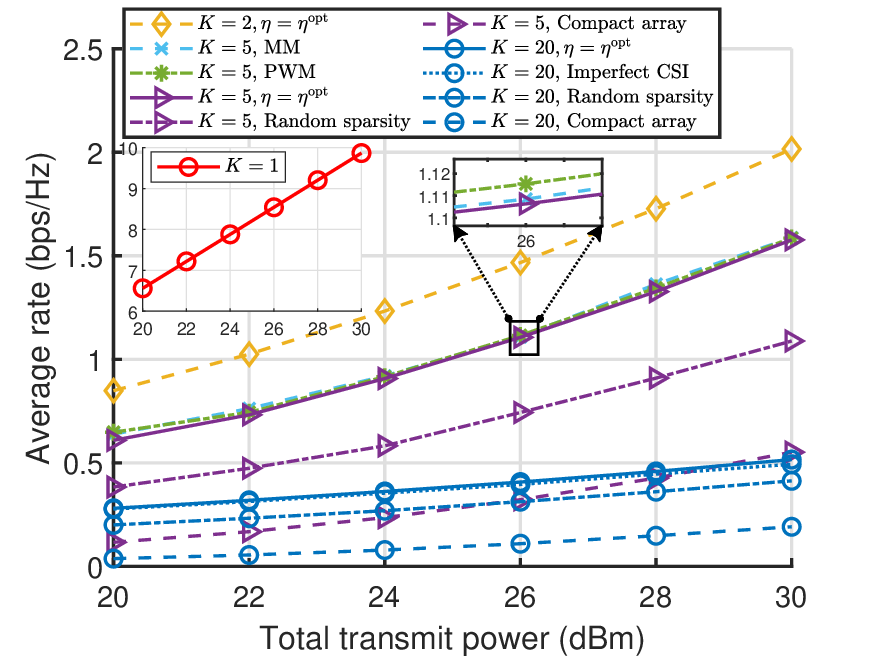}
  \vspace{-5pt}
 \caption{Average rate versus the total transmit power for different numbers of UEs.}
\label{fig: 4_rate_VS_K}
\vspace{-15pt}
\end{figure}

Fig. \ref{fig: 4_rate_VS_K} shows the average rate versus the total transmit power by considering different numbers of UEs. For comparison, the benchmark schemes PWM \cite{ji2025model} and MM \cite{zhang_RDARS} are considered.
It is observed that the average rate of the curve with $\eta^{\rm opt}$ when $K = 20$ surpasses that of the case with $\eta = 1$ when $K = 5$, thanks to the spatial gain brought by flexible sparsity designs.
Last but not least, the performance gap between the proposed sparsity design and benchmark schemes is negligible under the considered setup. In particular, the complexity of the proposed sparsity-based mode selection is $\mathcal{O}(\lfloor \frac{N-1}{a-1}\rfloor)$, rather than $\mathcal{O}(N^3)$ for the benchmark schemes. This demonstrates the effectiveness of the low-complexity sparsity design.
For the considered setup, the performance gap between the perfect CSI and imperfect CSI is relatively small, which demonstrates the robustness of proposed WA algorithm.

\section{Conclusion}
In this paper, we investigated a RDARS-aided system, where the sum rate was maximized by jointly optimizing the active beamforming of BS, passive beamforming of RDARS, and sparsity for RDARS CEA. To gain useful insights, the special cases of the single-UE and two-UE were respectively studied. Next, for an arbitrary number of UEs, the WA algorithm was proposed to jointly design the beamforming and sparsity level.
Numerical results showed that the proposed low-complexity sparsity design can achieve the comparable performance to the complex optimization algorithms. Last but not least, since the transmit power of connected element depends on the specific position when the fiber loss is considered, the power-dependent attenuation model is worthy of in-depth study in the future.



\bibliographystyle{IEEEtran}
\bibliography{IEEEabrv, Reference}
\end{document}